\documentclass[a4paper,11pt]{article}
\usepackage{pos}
\usepackage[english]{babel}
\title{Two-loop mixed QCD-EW corrections to neutral current Drell-Yan}

\author*[a]{Simone Devoto}

\affiliation[a]{Dipartimento di Fisica Aldo Pontremoli, Universit\`a di Milano and INFN, Sezione di Milano\\Via Celoria 16, I-20133 Milano, Italy}

\emailAdd{simone.devoto@unimi.it}

\abstract{We present the mixed QCD-EW two-loop virtual amplitudes for the neutral current Drell-Yan production, one of the bottlenecks for the complete calculation of the NNLO mixed QCD-EW corrections. We present the computational details and the first steps towards their automation.
We describe the evaluation of all the relevant two-loop Feynman integrals using analytical and semi-analytical methods, the subtraction of the universal infrared singularities and present the numerical evaluation of the finite remainder.}

\FullConference{%
  Loops and Legs in Quantum Field Theory - LL2022,\\
  25-30 April, 2022\\
  Ettal, Germany
}

\begin{document}
\maketitle

\section{Introduction}
The production of a lepton-pair with high transverse momentum, also known as Drell-Yan (DY) process, is of primary importance for precision programmes at hadron colliders: it has a clean experimental signature and a high production rate, allowing for a precise extraction of fundamental electroweak parameters.
In particular, thanks to the large amount of high-quality data collected by the experiments at the LHC, the measurements of several observables of interest, as the mass of the W boson\cite{ATLAS:2017rzl} or the electro-weak mixing angle\cite{CMS:2018ktx}, have been obtained with a precision which is starting to become competitive with previous results from LEP, and that is expected to reach sub-permille level by the end of the high-luminosity phase of the LHC.
Such experimental accuracy needs to be matched by precise theoretical Standard Model predictions, that also play a crucial role for new physics searches by providing severe constraints on possible models.

For these reasons, in the last years there has been an ongoing effort in order to improve the theoretical predictions on DY processes.
The computation of the on-shell Z boson production cross-section had some recent progress with the inclusion first of QCD-QED mixed corrections\cite{deFlorian:2018wcj,Delto:2019ewv,Hasan:2020vwn}, then mixed QCD-electroweak (EW) corrections\cite{Bonciani:2016wya,Bonciani:2019nuy,Bonciani:2020tvf,Bonciani:2021iis,Buccioni:2020cfi}.
Similar mixed corrections have also been computed for the production of an on-shell W boson\cite{Behring:2020cqi,Behring:2021adr}.

In the following, we will focus on the neutral-current DY process, where the final-state lepton pair is mediated by an off-shell photon or Z-boson:
\begin{equation}
\label{eq:process}
q(p_1)+\bar q(p_2)\rightarrow l^-(p_3)+l^+(p_4)\,.
\end{equation}
By considering a perturbative expansion in the strong ($\alpha_S$) and electroweak ($\alpha$) coupling, we can write the cross-section for this process as:
\begin{equation}
  \label{eq:expansion}
  d\sigma=\sum_{i,j} \alpha_S^i \alpha^j d\sigma^{(i,j)}\,,
\end{equation}
where $d\sigma^{(0,0)}$ is the leading order contribution.

The dominant effect from higher order corrections comes from the QCD corrections $d\sigma^{(i,0)}$, which have been computed at next-to-leading order (NLO)\cite{Altarelli:1979ub}, next-to-next-to-leading order (NNLO) \cite{Hamberg:1990np,Anastasiou:2003yy,Catani:2009sm} and, recently, up to next-to-next-to-next-to-leading order (N3LO)\cite{Ahmed:2014cla,Duhr:2020sdp,Chen:2021vtu,Camarda:2021ict,Chen:2022cgv,Neumann:2022lft}.

Electroweak corrections $d\sigma^{(0,j)}$ have a smaller impact, as suggested by the so-called "physical counting" $\alpha_S\simeq\alpha^2$. Nevertheless, they are not negligible. They are known up to NLO\cite{Baur:2001ze,Dittmaier:2001ay,Baur:2004ig}, while for NNLO, only  the Sudakov high energy approximation is available\cite{Jantzen:2005az}.

The large size of both NLO QCD and NLO EW corrections suggests that also the mixed strong-electroweak corrections $d\sigma^{(1,1)}$ might have a sizeable impact, which, by physical counting, is expected to be comparable with N3LO QCD contributions.
Recent results from two independent computations\cite{Bonciani:2021zzf,Buccioni:2022kgy} show indeed an effect of $\sim 0.5\%$ with respect to the LO result.
In this proceeding, we will present some technical aspects of the latter results, namely the computation of the mixed QCD-EW two-loop virtual corrections\cite{Armadillo:2022bgm} that represented one of the bottlenecks of the full calculation and that have been used to obtain the phenomenological results presented in\cite{Bonciani:2021zzf}.

\section{Computational framework}
The results presented in\cite{Bonciani:2021zzf} have been obtained by using the $q_T$-subtraction formalism\cite{Catani:2007vq} to treat and cancel singularities of infrared (IR) origin.
In the following, the cancellation of the IR poles of the virtual corrections is thus performed within this framework; our results can nevertheless be straightforwardly generalised to any other subtraction scheme by properly replacing the subtraction operator.

The $q_T$-subtraction formalism is at the moment only developed for the case of massive final-state emitters\cite{Bonciani:2015sha,Catani:2019iny}.
As a consequence, we keep in our computation the dependence on the lepton mass $m_l$ to regularise the final-state collinear singularities, while dropping it in the finite contributions.
We thus perform a small lepton mass limit, by considering the ratio $m_l/\sqrt{s}$ and by keeping only logarithmic terms $\simeq \log(m_l/\sqrt{s})$.

When dealing with intermediate unstable particles, such as the W or Z boson, it is useful to perform the calculation in the complex mass scheme in order to regularise the behaviour at the resonance.
In our computation we thus introduce a complex mass $\mu_V$ for the gauge boson $V= Z,W$, defined as:
\begin{equation}
  \mu_V^2=m_V^2-i\Gamma_V m_V\,,
\end{equation}
where the real parameters $m_V$ and $\Gamma_V$ are, respectively, the mass and the decay width of the gauge boson.
The introduction of the complex mass scheme also affects the kinematical variables of the process.
We define the Mandelstam variables:
\begin{equation}
  s=(p_1+p_2)^2\,, ~~ t=(p_1-p_3)^2\,,
\end{equation}
and their respective dimensionless kinematic invariants:
\begin{equation}
  \label{eq:complex_var}
  x_V=-\frac{s}{m_V^2}\,, ~~ y_V=-\frac{t}{m_V^2}\,.
\end{equation}
When replacing the pole mass with the complex mass, the adimensional variables $x_V$, $y_V$ become, in general, complex-valued.
As it will be shown in the following, this feature will require some additional care when dealing with the evaluation of the master integrals, in order to perform properly the analytic continuation of the solution in the complex plane.

\section{Evaluation of the interference term}
By following the expansion in Eq.(\ref{eq:expansion}), we can write the amplitude of the partonic process in Eq.(\ref{eq:process}) as:
\begin{equation}
   |{\cal M} \rangle = |{\cal M}^{(0)} \rangle + \alpha_S |{\cal M}^{(1,0)} \rangle + \alpha |{\cal M}^{(0,1)} \rangle + \alpha_S \alpha |{\cal M}^{(1,1)} \rangle + \cdots
\end{equation}
In order to evaluate the two-loop mixed QCD-EW corrections, we need to compute the following interference terms:
\begin{equation}
  \langle {\cal M}^{(0)} | {\cal M}^{(1,0)} \rangle \,, ~~
  \langle {\cal M}^{(0)} | {\cal M}^{(0,1)} \rangle \,, ~~
  \langle {\cal M}^{(0)} | {\cal M}^{(1,1)} \rangle \,.
  \label{eq:interferences}
\end{equation}

The first step is the generation of the relevant Feynman diagrams.
We used two completely independent approaches, one based on \textsc{FeynArts}\cite{Hahn:2000kx}, one based on \textsc{QGRAF}\cite{Nogueira:1991ex}.
Two independent in-house routines have then been used to automatically perform the Dirac and Lorentz Algebra.

The computation is performed in dimensional regularisation, which leads to the problem of handling consistently the inherently four-dimensional object $\gamma_5$ in $d=4-2\epsilon$ dimensions.
The prescription of 't Hooft and Veltman\cite{tHooft:1972tcz} proposes to abandon the anticommutation relation
\begin{equation}
  \{\gamma_\mu,\gamma_5\}=0\,,
\end{equation}
while keeping the cyclicity of the trace.
The prescription of Kreimer et al.\cite{Korner:1991sx}, on the other hand, suggests renouncing the cyclicity of the trace while keeping the anticommutation relation, reducing the computational load in a significative way.
It has been recently proven for neutral-current DY that at two loops the two prescription, while yielding to different scattering amplitudes, provide the same finite corrections after consistent subtraction of the IR and UV poles\cite{Heller:2020owb}.

In our computation, we keep the anticommutation relation of $\gamma_5$, using a fixed point to write the Dirac traces.
By using this propriety, we bring all the $\gamma_5$ matrices at the end of the Dirac trace, and by using the relation $\gamma_5^2=1$ we obtain a trace with, at most, a single leftover $\gamma_5$.
In the latter case, we use the identity
\begin{equation}
  \gamma_5=\frac i4 \epsilon_{\mu\nu \rho\sigma}\gamma^\mu\gamma^\nu\gamma^\rho\gamma^\sigma\;.
\end{equation}

At this stage, the interference terms are written as a sum of tensor integrals.
This expression, after some simple algebra, can be converted in terms of a sum of scalar integrals, expressed as elements of an integral family, each with the respective rational coefficient.
All the scalar integrals that appear in the expression, however, are not independent, and linear relations between them are provided by integration by parts (IBP) identities, that allow to reduce the original large set of scalar integrals to a smaller set of Master Integrals (MIs).

We executed the reduction to MIs by using two different public codes that implement Laporta algorithm
\cite{Laporta:2000dsw}
, \textsc{Kira}\cite{Maierhofer:2017gsa} and \textsc{LiteRed}\cite{Lee:2013mka}.
Our final basis of MI is composed by different integrals already known in the literature: MIs relevant for the QCD-QED corrections with massive final state\cite{Bonciani:2008az,Bonciani:2009nb}; MIs with one or two internal masses, relevant for the EW form factor\cite{Aglietti:2003yc,Aglietti:2004tq} and, finally, 31 MIs with 1 mass and 36 MIs with 2 masses (including boxes)\cite{Bonciani:2016ypc}, relevant for the QCD-EW corrections to the full DY.

\section{Semi-analytical solution of the master integrals: \textsc{SeaSyde}}
Despite the fact that all the MIs needed to complete our calculation were already studied in the literature, 5 box integrals with two internal massive lines were available\footnote{A closed form for them has been recently found\cite{Heller:2019gkq}, but is not yet public.} as Chen iterated integrals.
The difficult numerical evaluation of these functions requires finding alternative strategies.

We solved the 5 remaining MIs by using a semi-analytical approach.
We define a result as semi-analytical when it can be expanded as a power series at every point of its domain, but without the additional functional relations that are usually known when the result is provided in closed form.
In our computation, in particular, we express the 5 missing MIs as a Laurent expansion, which is obtained by solving by series the system of differential equations satisfied by the MIs\cite{Moriello:2019yhu}.

This algorithm has been implemented for real values of the kinematical variables in the \textsc{Mathematica} package \textsc{DiffExp}\cite{Hidding:2020ytt}\footnote{For a recent application of the same algorithm to the auxiliary mass flow method see also  \textsc{AMFlow}\cite{Liu:2022chg}.}.
Nevertheless, in our computation we needed to deal with complex-valued kinematical variables because of the introduction of the complex-mass scheme, as shown in Eq.(\ref{eq:complex_var}).
For this reason, we implemented the same method in an independent public \textsc{Mathematica} package, \textsc{SeaSyde}\cite{Armadillo:2022ugh}, generalising it in order to perform the analytic continuation of the solution on the complex plane.

Given a generic system of differential equations, we introduce an ansatz for the solution of the associated homogenous equation written in terms of a Laurent series expanded around the initial boundary condition $z_0$: $f_{\text{hom}}(z)=(z-z_0)^r \sum_{k=0}^\infty c_k (z-z_0)^k$.
The coefficients $c_k$ can be determined by plugging $f_{\text{hom}}$ in the homogenous system and by solving the set of algebraic equations obtained up to the desired order in the expansion.
This provides a homogenous solution, that can in turn be used to compute the particular solution for the original problem by using the variation of constant method.

\begin{figure}
  \centering
\includegraphics[width=0.4\textwidth]{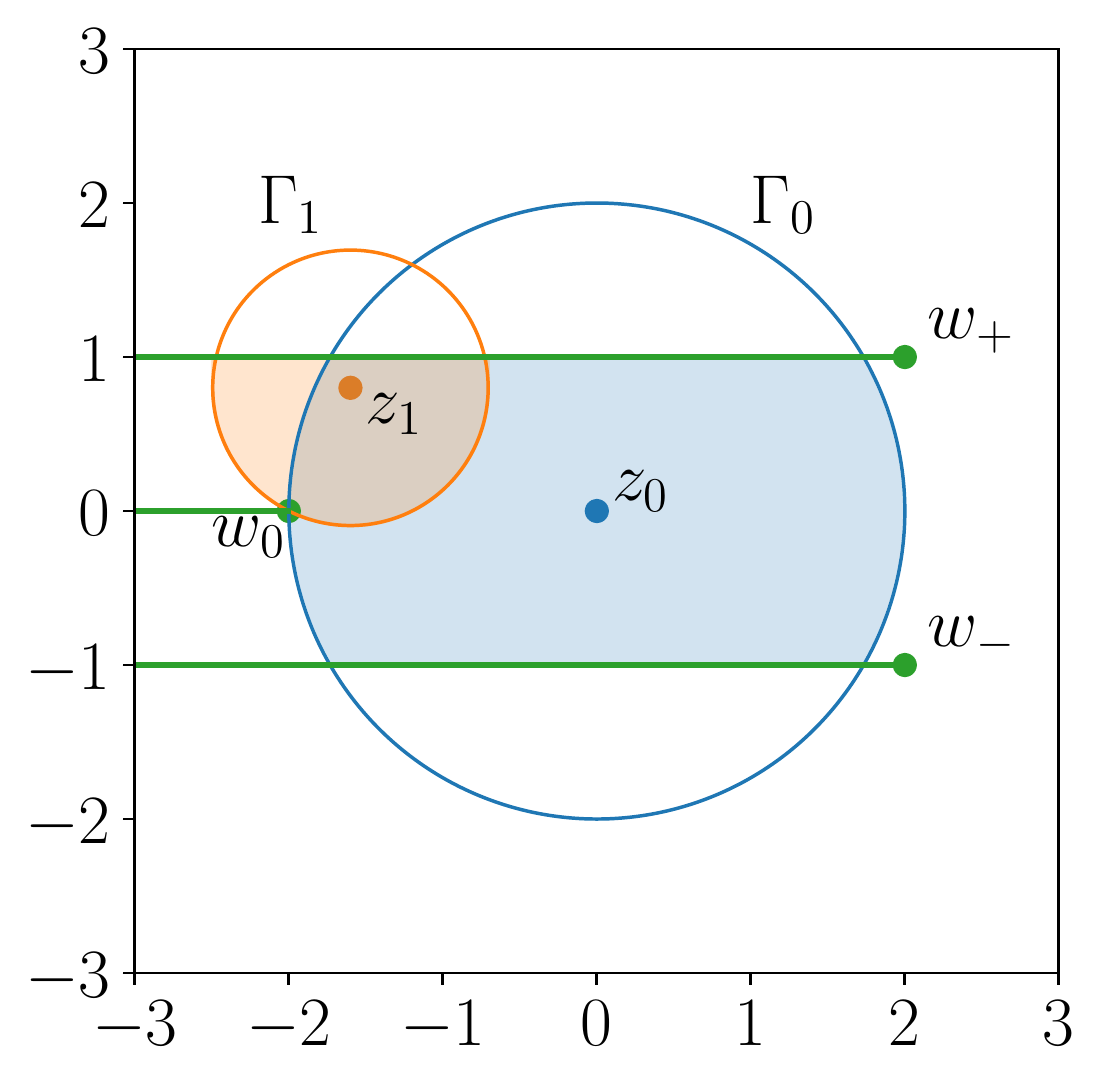}
\includegraphics[width=0.4\textwidth]{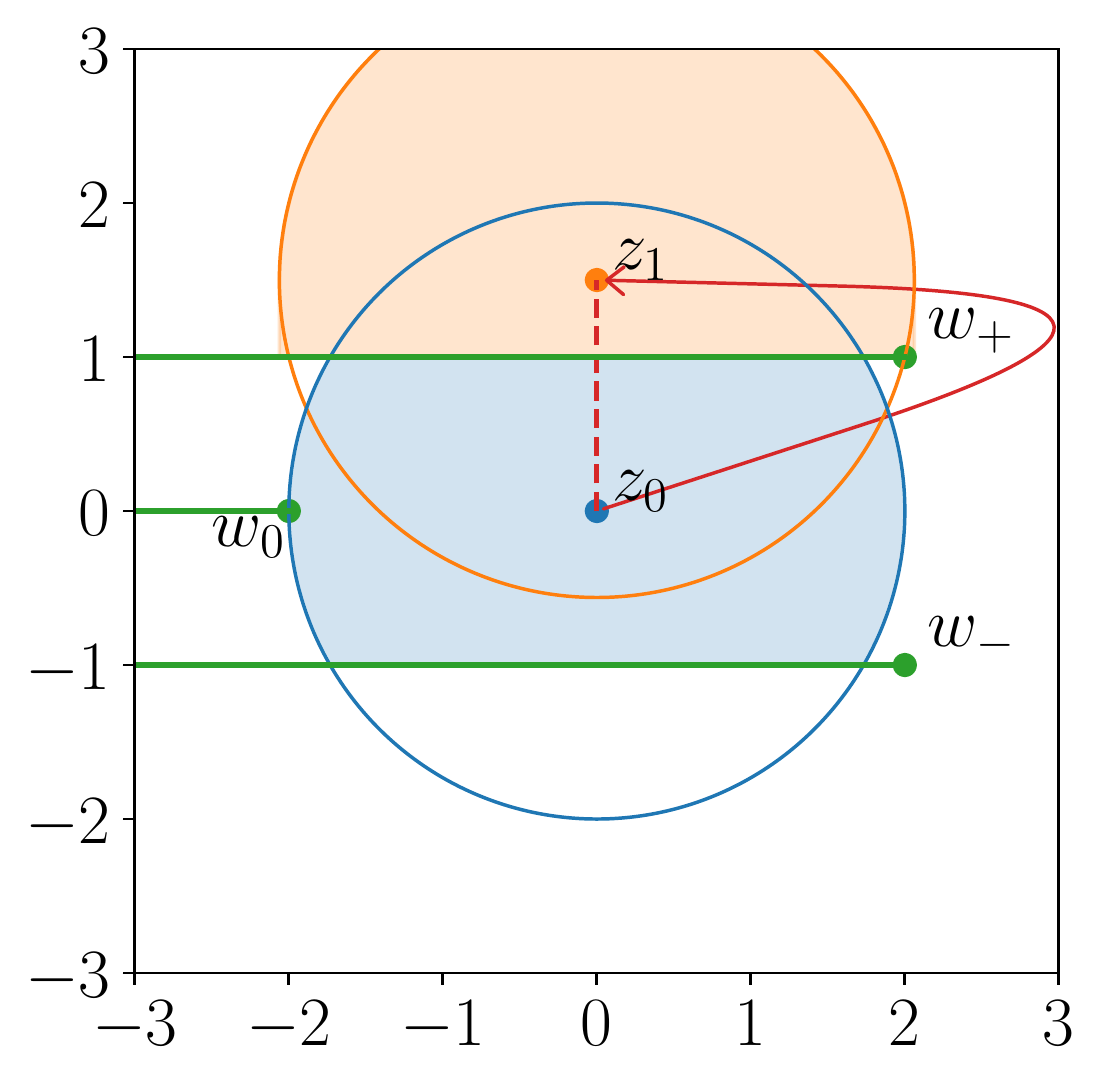}
\caption{\label{fig:cuts}
 Example of the effect of branch-cuts on the convergence of the expanded solution: reduced convergence area (left) and different path for the analytic continuation (right).}
\end{figure}

The solution obtained can be computed with an arbitrary number of significant digits, limited only by the precision of the boundary conditions.
It is valid within a radius of convergence given by the distance from the closest singular point, that can be directly obtained from the system of differential equations.
If this constraint does not allow evaluating the solution for the desired values of the kinematical variables, the procedure can be repeated using as a new boundary condition one of the points inside the radius of convergence.
With this procedure, the boundary condition can effectively be transported to any point of the complex plane.
This is illustrated in the left panel of Fig.\ref{fig:cuts}, where the pole $w_0$ limits the convergence of the solution expanded around $z_0$ within the circle $\Gamma_0$: nevertheless, the point $z_1$ can now be used as a new boundary condition to obtain the solution within the new circle $\Gamma_1$.

Some additional complications arise from the fact that, if the poles present a logarithmic behaviour, we need to insert branch-cuts to make the solution single-valued.
Within \textsc{SeaSyde}, the branch-cuts are always chosen as the horizontal lines parallel to the real axis that go from the singular point to $-\infty$.
While their presence does not affect the radius of convergence, it reduces the area in which the solution converges to the desired value, as shown in the left panel of Fig.\ref{fig:cuts}: once the branch-cut is crossed, the solution converges to a value that does not refer anymore to the Riemann sheet which is consistent with the branch-cut itself.
For the same reason, the path chosen to transport the boundary condition from one point to another requires to avoid to cross the branch cuts: this is shown in the right panel of Fig.\ref{fig:cuts}, where the dotted path needs to be avoided in favour of the solid path.

\section{Results and conclusions}
We used the package \textsc{SeaSyde} to solve the system of differential equations associated to the 36 MIs with 2 internal masses.
The result of 31 MIs provided a cross check with the known analytic expressions, while 5 MIs, the ones known as Chen iterated integrals, are a prediction.
Several checks on the MIs have been performed by using \textsc{Fiesta}\cite{Smirnov:2015mct}, \textsc{PySecDec}\cite{Borowka:2017idc} and \textsc{DiffExp}.

By combining the rational coefficients with the expression of the MIs, after the subtraction of the infrared and ultraviolet divergences, we obtained the two-loop virtual corrections for neutral-current DY process in the complex-mass scheme and in the small lepton mass limit, keeping the collinear logarithms.

The result is publicly available as a \textsc{Mathematica} notebook\cite{Armadillo:2022bgm} in the form of a grid.
The production of the grid required $\mathcal O(12\text{h})$ on a 32-cores machine, but the interpolation of the grid can be performed in negligible time.
While phenomenological results obtained by using this computation have been already presented\cite{Bonciani:2021zzf}, more detailed studies are ongoing.
Furthermore, the automatic nature of several steps of the procedure outlined in this proceeding leaves the door open to several further applications, including mixed corrections for charged current DY\footnote{A first result for mixed correction to charged DY process has been already presented\cite{Buonocore:2021rxx}, where the 2-loop contributions are at the moment expressed in pole approximation.} and, possibly, first steps towards NNLO-EW corrections.

\section*{Acknowledgements}
This proceeding is based upon work done in collaboration with T. Armadillo, R. Bonciani, N. Rana, A. Vicini and supported by the Italian Ministero della Universit\'a e della Ricerca (grant PRIN201719AVICI 01).

\bibliographystyle{JHEP}
\bibliography{biblio}
\end{document}